\documentclass[twocolumn]{aastex631}

\graphicspath{{./}}

\received{}
\revised{}
\accepted{}
\submitjournal{ApJ Letters}

\shorttitle{FRB\,20220912A localization}
\shortauthors{Ravi et al.}

\begin{document}

\title{Deep Synoptic Array science I: discovery of the host galaxy of FRB\,20220912A}

\correspondingauthor{Vikram Ravi}
\email{vikram@caltech.edu}

\author{Vikram Ravi}
\affiliation{Cahill Center for Astronomy and Astrophysics, MC 249-17 California Institute of Technology, Pasadena CA 91125, USA.}
\affiliation{Owens Valley Radio Observatory, California Institute of Technology, Big Pine CA 93513, USA.}

\author{Morgan Catha}
\affiliation{Owens Valley Radio Observatory, California Institute of Technology, Big Pine CA 93513, USA.}

\author{Ge Chen}
\affiliation{Cahill Center for Astronomy and Astrophysics, MC 249-17 California Institute of Technology, Pasadena CA 91125, USA.}

\author{Liam Connor}
\affiliation{Cahill Center for Astronomy and Astrophysics, MC 249-17 California Institute of Technology, Pasadena CA 91125, USA.}

\author{Jakob T. Faber}
\affiliation{Cahill Center for Astronomy and Astrophysics, MC 249-17 California Institute of Technology, Pasadena CA 91125, USA.}

\author{James W. Lamb}
\affiliation{Owens Valley Radio Observatory, California Institute of Technology, Big Pine CA 93513, USA.}

\author{Gregg Hallinan}
\affiliation{Cahill Center for Astronomy and Astrophysics, MC 249-17 California Institute of Technology, Pasadena CA 91125, USA.}
\affiliation{Owens Valley Radio Observatory, California Institute of Technology, Big Pine CA 93513, USA.}

\author{Charlie Harnach}
\affiliation{Owens Valley Radio Observatory, California Institute of Technology, Big Pine CA 93513, USA.}

\author{Greg Hellbourg}
\affiliation{Cahill Center for Astronomy and Astrophysics, MC 249-17 California Institute of Technology, Pasadena CA 91125, USA.}
\affiliation{Owens Valley Radio Observatory, California Institute of Technology, Big Pine CA 93513, USA.}

\author{Rick Hobbs}
\affiliation{Owens Valley Radio Observatory, California Institute of Technology, Big Pine CA 93513, USA.}

\author{David Hodge}
\affiliation{Cahill Center for Astronomy and Astrophysics, MC 249-17 California Institute of Technology, Pasadena CA 91125, USA.}

\author{Mark Hodges}
\affiliation{Owens Valley Radio Observatory, California Institute of Technology, Big Pine CA 93513, USA.}

\author{Casey Law}
\affiliation{Cahill Center for Astronomy and Astrophysics, MC 249-17 California Institute of Technology, Pasadena CA 91125, USA.}
\affiliation{Owens Valley Radio Observatory, California Institute of Technology, Big Pine CA 93513, USA.}

\author{Paul Rasmussen}
\affiliation{Owens Valley Radio Observatory, California Institute of Technology, Big Pine CA 93513, USA.}

\author{Kritti Sharma}
\affiliation{Cahill Center for Astronomy and Astrophysics, MC 249-17 California Institute of Technology, Pasadena CA 91125, USA.}

\author{Myles B. Sherman}
\affiliation{Cahill Center for Astronomy and Astrophysics, MC 249-17 California Institute of Technology, Pasadena CA 91125, USA.}

\author{Jun Shi}
\affiliation{Cahill Center for Astronomy and Astrophysics, MC 249-17 California Institute of Technology, Pasadena CA 91125, USA.}

\author{Dana Simard}
\affiliation{Cahill Center for Astronomy and Astrophysics, MC 249-17 California Institute of Technology, Pasadena CA 91125, USA.}

\author{Reynier Squillace}
\affiliation{Cahill Center for Astronomy and Astrophysics, MC 249-17 California Institute of Technology, Pasadena CA 91125, USA.}

\author{Sander Weinreb}
\affiliation{Cahill Center for Astronomy and Astrophysics, MC 249-17 California Institute of Technology, Pasadena CA 91125, USA.}

\author{David P. Woody}
\affiliation{Owens Valley Radio Observatory, California Institute of Technology, Big Pine CA 93513, USA.}

\author{Nitika Yadlapalli}
\affiliation{Cahill Center for Astronomy and Astrophysics, MC 249-17 California Institute of Technology, Pasadena CA 91125, USA.}

\collaboration{200}{(The Deep Synoptic Array team)}

\author{Tomas Ahumada}
\affiliation{Cahill Center for Astronomy and Astrophysics, MC 249-17 California Institute of Technology, Pasadena CA 91125, USA.}

\author{Dillon Dong}
\affiliation{Cahill Center for Astronomy and Astrophysics, MC 249-17 California Institute of Technology, Pasadena CA 91125, USA.}

\author{Christoffer Fremling}
\affiliation{Cahill Center for Astronomy and Astrophysics, MC 249-17 California Institute of Technology, Pasadena CA 91125, USA.}

\author{Yuping Huang}
\affiliation{Cahill Center for Astronomy and Astrophysics, MC 249-17 California Institute of Technology, Pasadena CA 91125, USA.}

\author{Viraj Karambelkar}
\affiliation{Cahill Center for Astronomy and Astrophysics, MC 249-17 California Institute of Technology, Pasadena CA 91125, USA.}

\author{Jessie M. Miller}
\affiliation{Cahill Center for Astronomy and Astrophysics, MC 249-17 California Institute of Technology, Pasadena CA 91125, USA.}

\begin{abstract}

We report the detection and interferometric localization of the repeating fast radio burst (FRB) source FRB\,20220912A during commissioning observations with the Deep Synoptic Array (DSA-110). Two bursts were detected from FRB\,20220912A, one each on 2022 October 18 and 2022 October 25. The best-fit position is (R.A. J2000, decl. J2000) = (23:09:04.9, $+$48:42:25.4), with a 90\% confidence error ellipse of $\pm2$\arcsec~ and $\pm1$\arcsec~ in right ascension and declination respectively. The two bursts have disparate polarization properties and temporal profiles. We find a Faraday rotation measure that is consistent with the low value of $+0.6$\,rad\,m$^{-2}$ reported by CHIME/FRB. The DSA-110 localization overlaps with the galaxy PSO\,J347.2702$+$48.7066 at a redshift $z=0.0771$, which we identify as the likely host. PSO\,J347.2702$+$48.7066 has a stellar mass of approximately $10^{10}M_{\odot}$, modest internal dust extinction, and a star-formation rate likely in excess of $0.1\,M_{\odot}$\,yr$^{-1}$. The host-galaxy contribution to the dispersion measure is likely $\lesssim50$\,pc\,cm$^{-3}$. The FRB\,20220912A source is therefore likely viewed along a tenuous plasma column through the host galaxy. 

\end{abstract}


\section{Introduction} \label{sec:intro}

Eight repeating sources of fast radio bursts (FRBs) have been localized to host galaxies \citep{clw+17,mnh+20,mpm+20,bkm+21,kmn+22,rll+22,bha+22,nal+22}. It is not yet established whether there are astrophysical differences between the progenitors of the few FRB sources that have been observed to repeat, and the many FRB sources that have not yet been observed to repeat. Although it is likely that most FRB sources repeat at some level \citep{r19}, repeating sources are different from apparent non-repeaters in their burst morphologies and bandwidths \citep{pgk+21}, and in their repetition rates \citep{jof+20}. No significant differences between the host galaxies of repeaters and apparent non-repeaters are yet evident \citep{bha+22}. The association of persistent radio sources with some repeaters may hint at a unique progenitor class  \citep{lca+22}. 

Repeating FRB sources offer the best means to identify the objects and mechanisms that can produce radio bursts with energy outputs in excess of $\sim10^{35}$\,erg on $\ll1$\,s timescales. This is particularly the case when observations of repeaters are placed in the rich context of milliarcsecond-scale localizations. For example, the association of FRB\,20121102A with a star-forming region and compact persistent radio source within its dwarf host \citep{bta+17}, combined with its extreme and variable Faraday rotation measure \citep[RM;][]{msh+18}, are suggestive of a young compact object within its birth supernova remnant \citep[][and references therein]{crh22}. The detection of microstructure in bursts from FRB\,20200120E \citep{mpp+21,nhk+22}, together with its association with a globular cluster of M81 \citep{kmn+22}, suggest an origin in a recycled pulsar system. Despite its origin in an inter-arm region of its late-type host galaxy \citep{xnc+22}, FRB\,20201124A illuminates extreme variability of its sub-AU magneto-ionic environment, suggestive of a magnetar/Be-star binary \citep{wzd+22}. Such observational diversity among the small sample of just eight localized sources motivates the characterization of a larger sample of host galaxies and environments of repeating FRBs. 

FRB\,20220912A is an intensely active repeating FRB source discovered by the CHIME/FRB collaboration \citep{mc22}. 12 bursts were detected in the 400--800\,MHz band between 2022 September 12 and 2022 October 15. Throughout this paper, unless otherwise indicated, we adopt the dispersion measure (DM) and RM derived by CHIME/FRB from the brightest reported burst, 219.46\,pc\,cm$^{-3}$ and $+0.6$\,rad\,m$^{-2}$ respectively. On 2022 October 18, \citet{h22} reported the detection of 12 bursts within 49\,hr of exposure at teh Stockert telescope, with fluences $>10$\,Jy\,ms in the 1330--1430\,MHz band. A burst rate in excess of 400\,hr$^{-1}$ was found in the 1000--1500\,MHz band in observations with the Five hundred meter Aperture Spherical Telescope \citep[FAST;][]{znf+22}, and the Arecibo 12-m telescope was used to detect the source at  2.3\,GHz \citep{ppf+22}. Deep limits on transient optical emission coincident with several radio bursts from FRB\,20220912A were reported by \citet{hbm+22}.

Here we report the detection and interferometric localization of FRB\,20220912A during commissioning of the Deep Synoptic Array (DSA-110). Preliminary results from this work were reported via Astronomer's Telegram \citep{r+22a,r+22b,r+22c}. In \S\ref{sec:obs} we describe the DSA-110 (\S\ref{sec:dsa110}) and observations obtained of FRB\,20220912A (\S\ref{sec:radioobs}), and present the detections (\S\ref{sec:detect}) and localizations (\S\ref{sec:loc}) of two bursts. In \S\ref{sec:host} we present observations of the likely host galaxy of FRB\,20220912A, PSO\,J347.2702$+$48.7066, obtained at the W.~M.~Keck Observatory, together with modeling of fundamental properties of the host. We summarize and discuss implications of these results in \S\ref{sec:summary}. Throughout we adopt a flat cosmology with parameters derived by \citet{pc20}, including a Hubble constant of $67.4$\,km\,s$^{-1}$\,Mpc$^{-1}$, and a matter-density parameter of $\Omega_{m}=0.315$.

\section{DSA-110 observations} \label{sec:obs}

\subsection{Description of the DSA-110} \label{sec:dsa110}

The DSA-110\footnote{\url{https://deepsynoptic.org}} is a radio interferometer hosted at the Owens Valley Radio Observatory (OVRO). During the observations presented herein, the array was being commissioned with data from 63 antennas being processed. A full description will be presented in Ravi et al. (in prep.), and we here include essential information to support the observational results for FRB\,20220912A. 

\begin{deluxetable}{cc}
\tablecaption{Specifications of the DSA-110 during commissioning observations.\label{tab:dsa}}
\tablehead{
\colhead{Parameter} & \colhead{Value}}
\startdata
Dish diameter (m) & 4.65 \\
Central frequency (MHz) & 1405 \\
Primary beam FWHM (degrees) & 3.4 \\
Bandwidth (MHz) & 187.5 \\
Search SEFD (Jy)\tablenotemark{a} & 140 \\
Search-beam channel width (kHz) & 244.141 \\
Search-beam time resolution ($\mu$s) & 262.144 \\
Search-beam width (arcsec.)\tablenotemark{b} & 134 \\
Number of search beams & 256 \\
Search-beam spacing (arcmin.) & 1 \\ 
Maximum baseline (m) & 2500 \\
Synthesized beam\tablenotemark{c} & 35.4\arcsec$\times$17.6\arcsec @ $83.4^{\circ}$ 
\enddata
\tablenotetext{a}{System-equivalent flux density (SEFD) at boresight.}
\tablenotetext{b}{This is in the east-west direction. The search beams are fan beams formed with an east-west array, and their sizes are only constrained by the primary beam in the north-south direction.}
\tablenotetext{c}{Major and minor-axis diameters at a declination of $+48.7^{\circ}$, assuming natural weighting. The position angle is measured east of north.}
\end{deluxetable}

Essential specifications of the DSA-110 at the time of these observations are presented in Table~\ref{tab:dsa}. Each DSA-110 antenna is automatically steerable in elevation only, and observes on the meridian. Each antenna is equipped with dual orthogonal linearly polarized receivers and ambient temperature amplifiers that deliver a typical system temperature of 25\,K \citep{ws21}. A real-time FRB search is enabled through the coherent combination of 48 antennas located in an east-west line, with a maximum spacing of 400\,m. Total-intensity data in 256 coherent fan-shaped search beams are inspected in real time for FRBs, using a modified version of the \texttt{heimdall} software \citep{bbb+12}. Triggers from FRB candidates result in the storage of post-filterbank voltage data from the 48 search antennas and 15 outrigger antennas. The 4-bit voltage data are recorded for 61440 samples at 32.768\,$\mu$s time resolution in $6144\times30.518$\,kHz channels. 

Data from all antennas are also continuously correlated in real time in the two linear polarizations only (no cross-polarization products), resulting in the production of visibility data fringestopped on the meridian at the pointing center. Each FRB candidate is accompanied by the storage of 2\,hr of visibility data surrounding the trigger time. These visibility data are used to solve for antenna-based complex gains at the trigger time using a sky model. Any $>1$\,Jy calibrator source from the VLA calibrator manual that passes within the primary beam, represents $>15$\% of the flux within the beam, and is classified as `P' or `S' for the VLA C configuration, is also used to trigger the storage of 10\,min of visibility data fringestopped at the calibrator location. These visibility data are used to derive antenna-based bandpass calibrations for the beamformer, and for post-processing of the voltage data. 

\subsection{Observations of FRB\,20220912A} \label{sec:radioobs}

Following the report on FRB\,20220912A by CHIME/FRB \citep{mc22} on 2022 October 15, we began a campaign of near-daily observations of transits of the source with the DSA-110. The array was pointed at a declination of 48.7$^{\circ}$. We report on the period between 2022 October 15 and 2022 October 25, during which we observed seven transits of FRB\,20220912A and detected two bursts with the real-time system. During each transit the source was within the field of view of the search beams for 25.9\,min. As these observations were undertaken during science commissioning, and the precise search completeness was not quantified, we do not quote a derived burst rate from FRB\,20220912A. 

When not observing FRB\,20220912A, the DSA-110 was primarily pointed at a declination of 71.6$^{\circ}$. Daily bandpass calibration was made possible using observations of the source 3C309.1 (J1459$+$7140), which has a flux density of 7.6\,Jy at a wavelength of 20\,cm according to the VLA calibrator manual. Approximate polarization calibration was made possible using voltage data obtained on the standard sources 3C48 and 3C286 \citep{pb13} on 2022 October 20.  

\subsection{Two burst detections} \label{sec:detect}

\begin{figure*}
    \centering
    \includegraphics[width=0.4\textwidth]{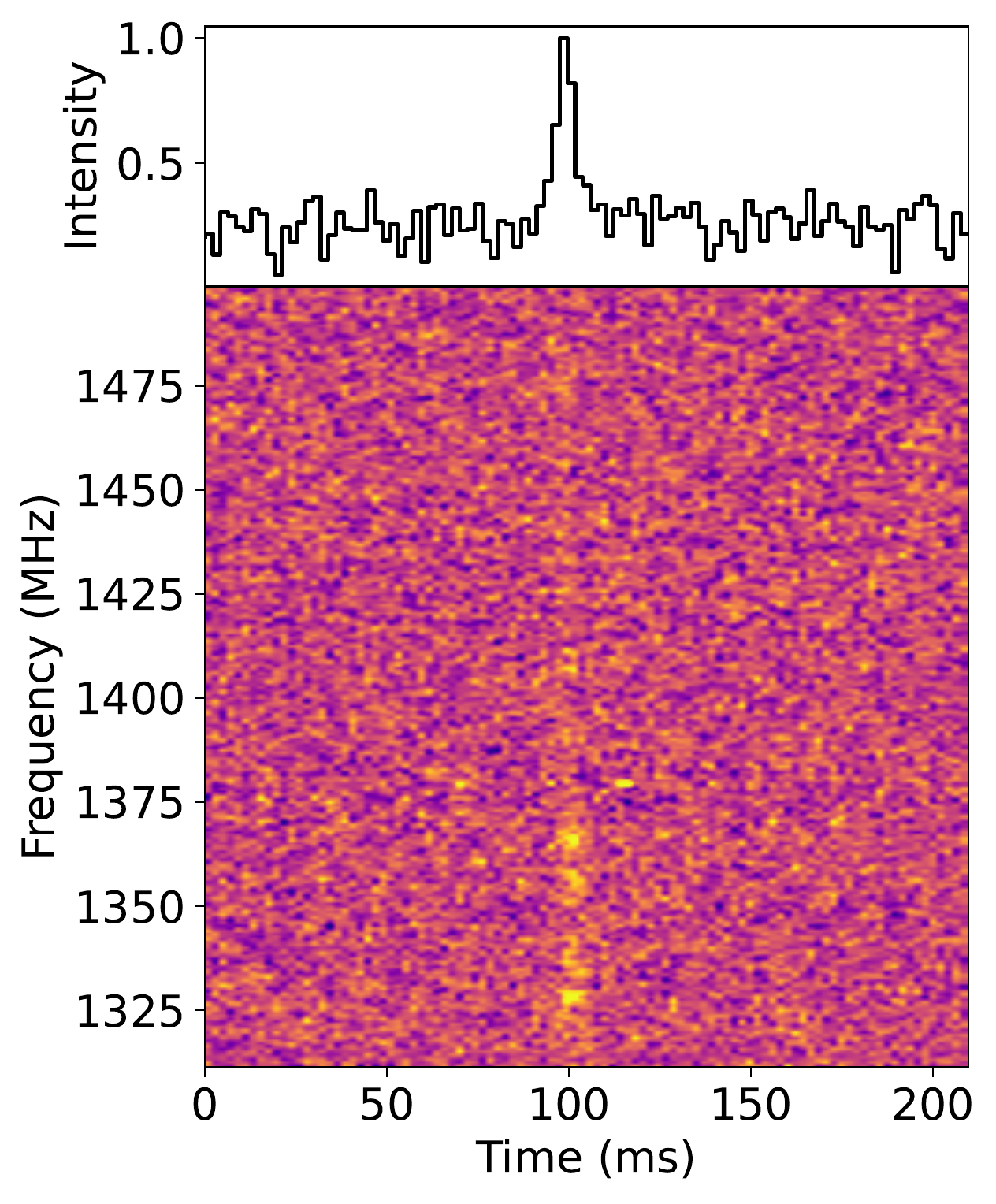}
    \includegraphics[width=0.4\textwidth]{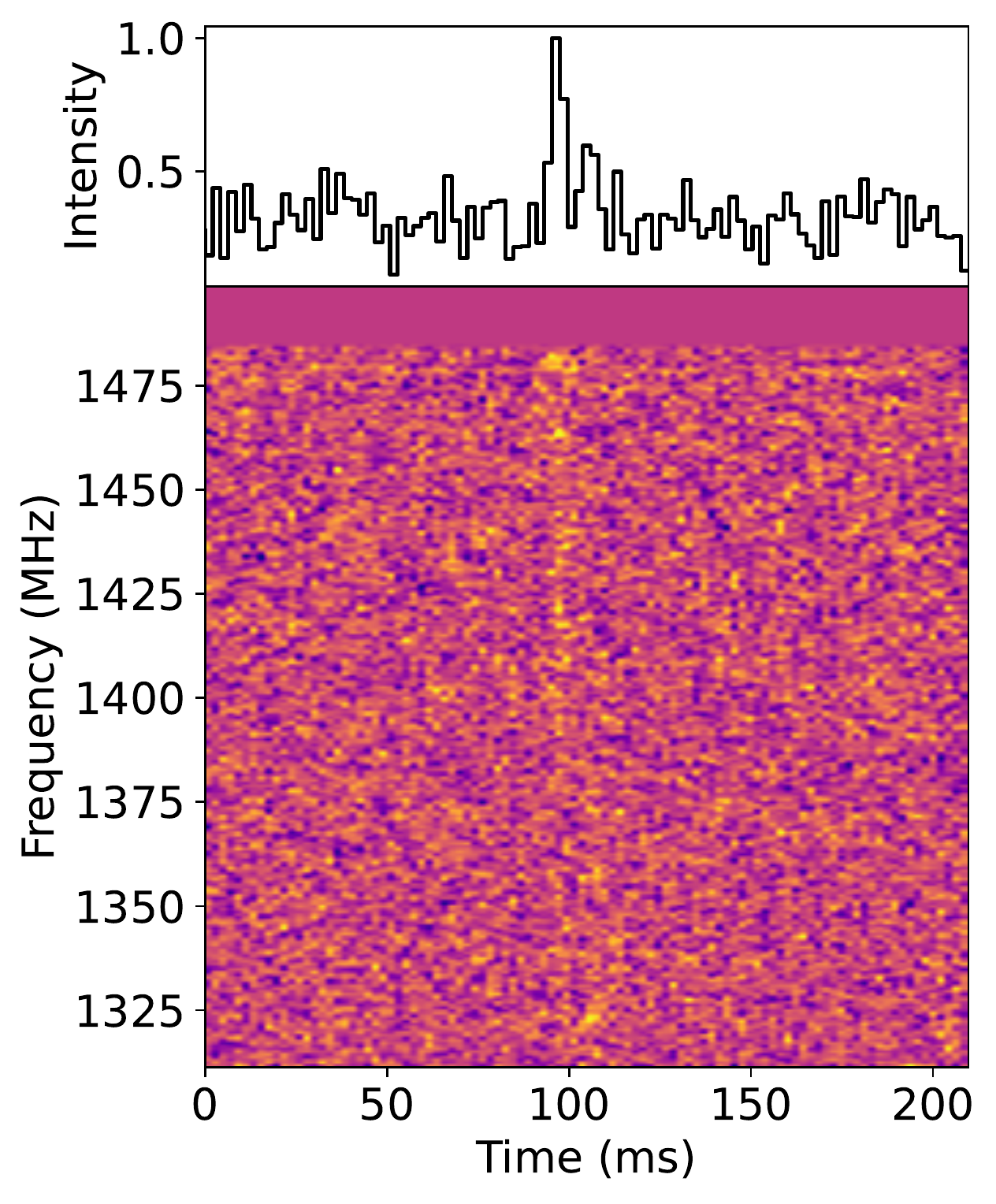}
    \caption{Dedispersed total-intensity dynamic spectra for bursts 1 (left) and 2 (right) detected with the DSA-110 from FRB\,20220912A. Both bursts have been dedispersed to ${\rm DM}=219.46$\,pc\,cm$^{-3}$. Voltage data in one out of 16 sub-bands for Burst 2 were corrupted, resulting in the blanked data at the top of the band.}
    \label{fig:bursts}
\end{figure*}

\begin{figure*}
    \centering
    \includegraphics[trim={4cm 2cm 4cm 2cm},clip,width=0.44\textwidth]{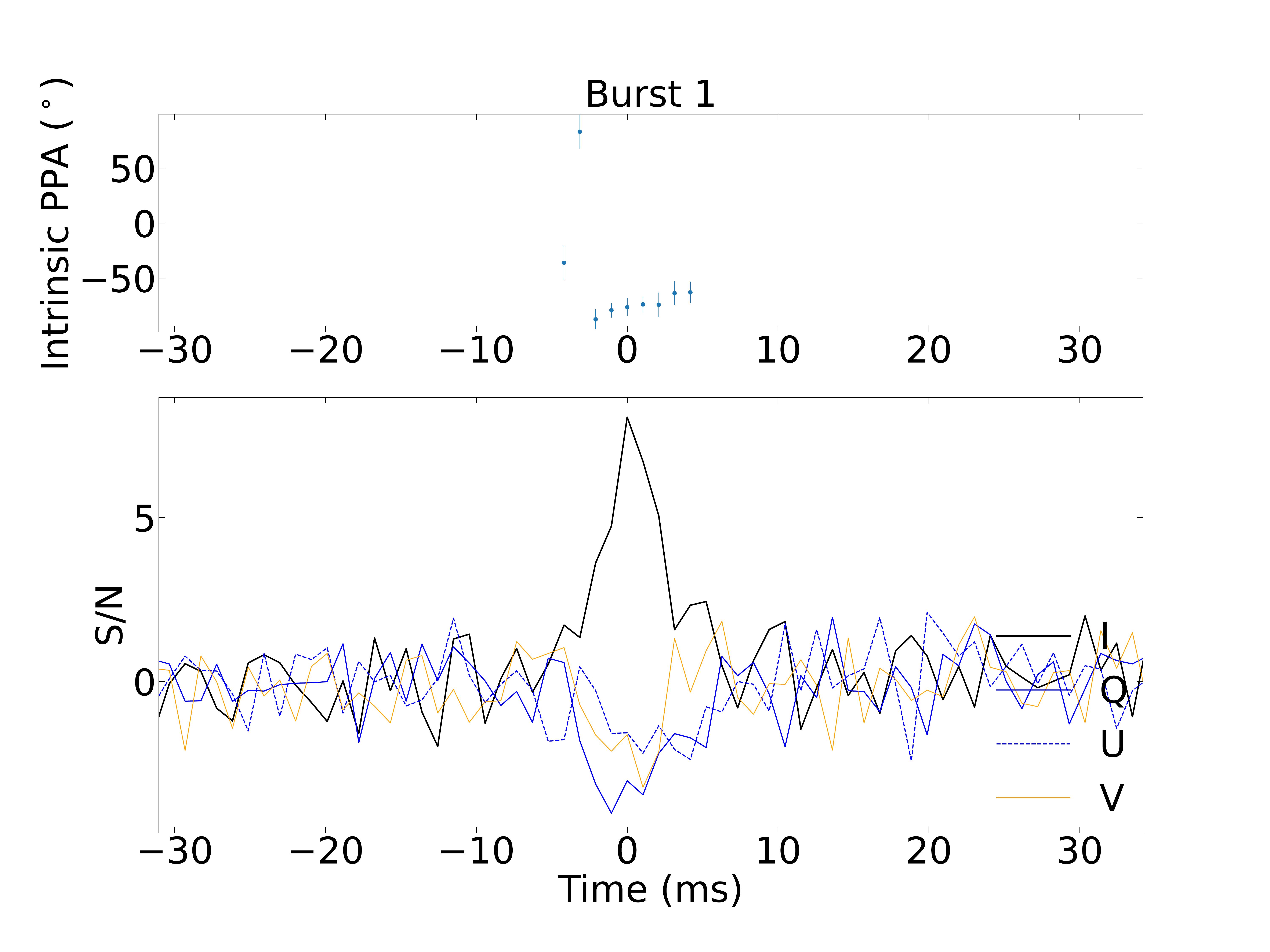}
    \includegraphics[trim={4cm 2cm 4cm 2cm},clip,width=0.44\textwidth]{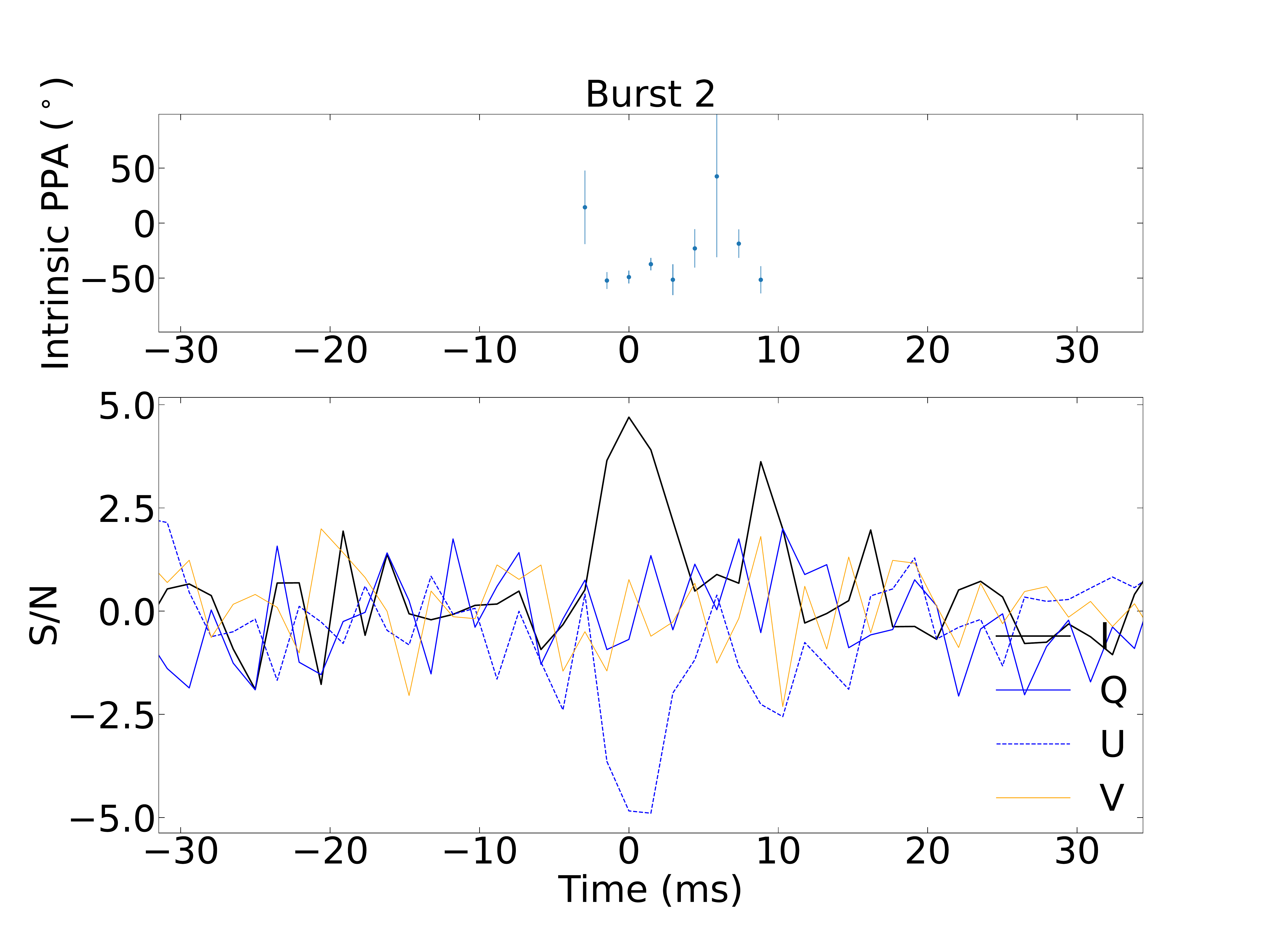}
    \caption{Polarization profiles of bursts 1 (left) and 2 (right) detected by the DSA-110 from FRB\,20220912A. Both bursts have been dedispersed to ${\rm DM}=219.46$\,pc\,cm$^{-3}$. The top panels show the absolute polarization position angle, and Stokes parameters are as labeled in the bottom panels. The polarized dynamic spectra of the bursts were corrected for the small RM reported by CHIME/FRB (0.6\,rad\,m$^{-2}$).}
    \label{fig:pol}
\end{figure*}

We detected bursts from FRB\,20220912A on 2022 October 18 (hereafter Burst 1) and on 2022 October 25 (hereafter Burst 2), on modified Julian dates (MJDs) 59870.22278235404 and 59877.203253573236 respectively. The arrival times correspond to 1530\,MHz and the DSA-110 reference position of $-118.283^{\circ}$ longitude, $+37.2334^{\circ}$ latitude. Burst 1 was detected with a signal to noise ratio (S/N) of 12.2 and an approximate DM of 228\,pc\,cm$^{-3}$, and Burst 2 was detected with an S/N of 8.5 and an approximate DM of 223\,pc\,cm$^{-3}$. Despite the differences between these DMs and the structure-optimized DM reported by CHIME/FRB \citep{mc22}, we associated these detections with FRB\,20220912A based on the correspondance between the CHIME/FRB position and the detection search-beam. We attribute these differences to the time-frequency drifting common in repeating FRBs \citep[e.g.,][]{hss+19}.

Voltage data on each burst were coherently combined in the direction of the detection beams using the 48 core antennas only. Time-frequency data sets in all four Stokes parameters were formed with the native time and frequency resolutions. Incoherent dedispersion at the native time and frequency resolutions was applied. Flattening of the bandpasses in each received polarization was done on a per-antenna basis using observations of 3C309.1 on the same sidereal days as the burst detections. Voltage data obtained during transits of 3C48 and 3C286 were used for polarization calibration under the ideal feed assumption; coupling between the orthogonal polarizations is detected at only the percent level for DSA-110 (Sherman et al., in prep.). Full-Stokes spectra towards 3C48 and 3C286 were formed using the same beamforming procedure as applied to the burst data. Spectra on 3C48 were used to verify the equalization of the amplitude gains in the two linear polarizations, and spectra on 3C286 were used to measure the frequency-dependent phase between the two polarizations. 

Total intensity dynamic spectra and temporal profiles are shown in Figure~\ref{fig:bursts}, and full-Stokes temporal profiles are shown in Figure~\ref{fig:pol}. Voltage data in one out of 16 sub-bands for Burst 2 were corrupted, resulting in the reduced bandwidth evident in Figure~\ref{fig:bursts}. The two bursts exhibit different morphologies and polarization characteristics. Burst 1 exhibits a single component whereas Burst 2 exhibits two components with visually different total-intensity spectra. For Burst 1, we measure a linear polarization fraction of $64\%\pm18\%$ and a circular polarization fraction of $41\%\pm9\%$. For Burst 2, we measure a linear polarization fraction of $73\%\pm30\%$, with no significant circular polarization, and a different polarization position angle. This is consistent with the report in \citet{znf+22} of substantially varying polarization properties between bursts from FRB\,20220912A. Although these data have been corrected for an RM of $+0.6$\,rad\,m$^{-2}$ \citep{mc22}, an independent RM search of each burst revealed measurements consistent with this value. These data do not have sufficiently high S/N for more detailed inferences to be made. 

\subsection{Interferometric localization} \label{sec:loc}

\begin{figure*}
    \centering
    \includegraphics[width=0.85\textwidth]{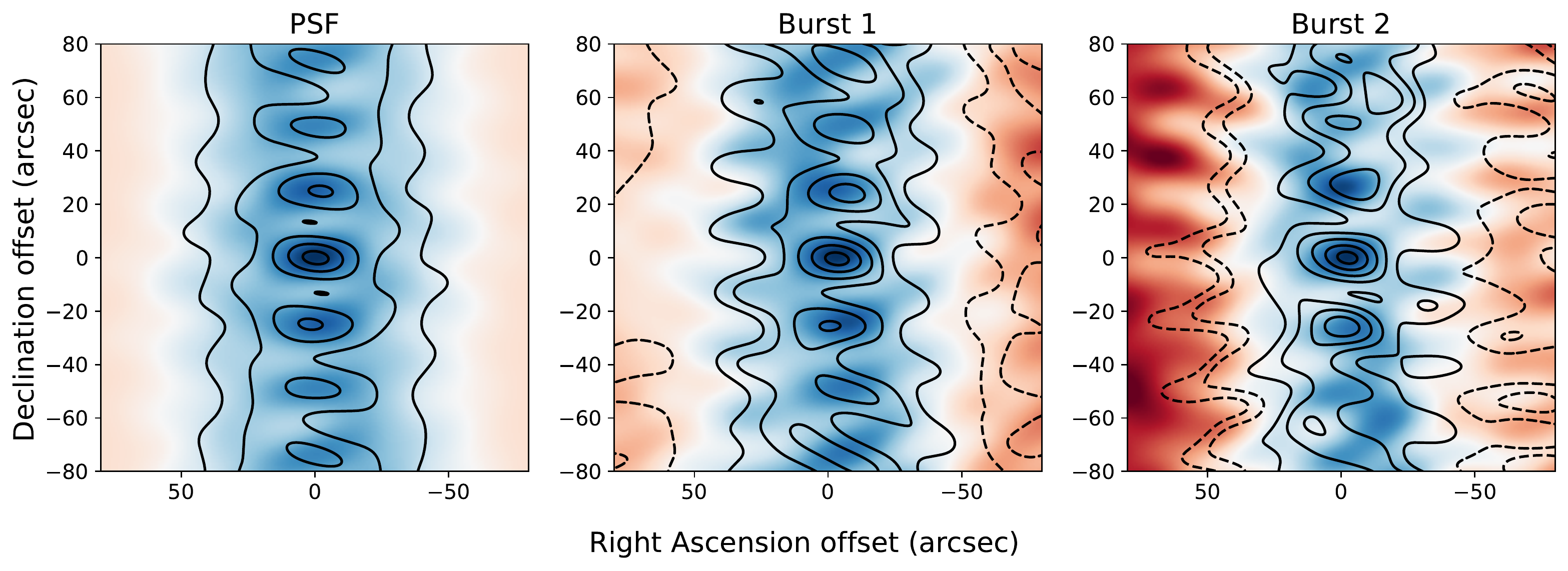}
    \caption{Images of Burst 1 (middle) and Burst 2 (right) from FRB\,20220912A, together with the DSA-110 PSF for Burst 1 (left). The PSF for Burst 2 is negligibly different from that for Burst 1. No deconvolution has been applied to the burst images. Contours are at $-0.4$, $-0.2$ (dashed), $0.2$, $0.4$, $0.6$, $0.8$ and $0.9$ (solid) of the peak intensity. The burst images are centered on the coordinates (R.A. J2000, decl. J2000) = (23:09:04.9, $+$48:42:25.4).}
    \label{fig:localizations}
\end{figure*}

\begin{figure}
    \centering
    \includegraphics[width=0.45\textwidth]{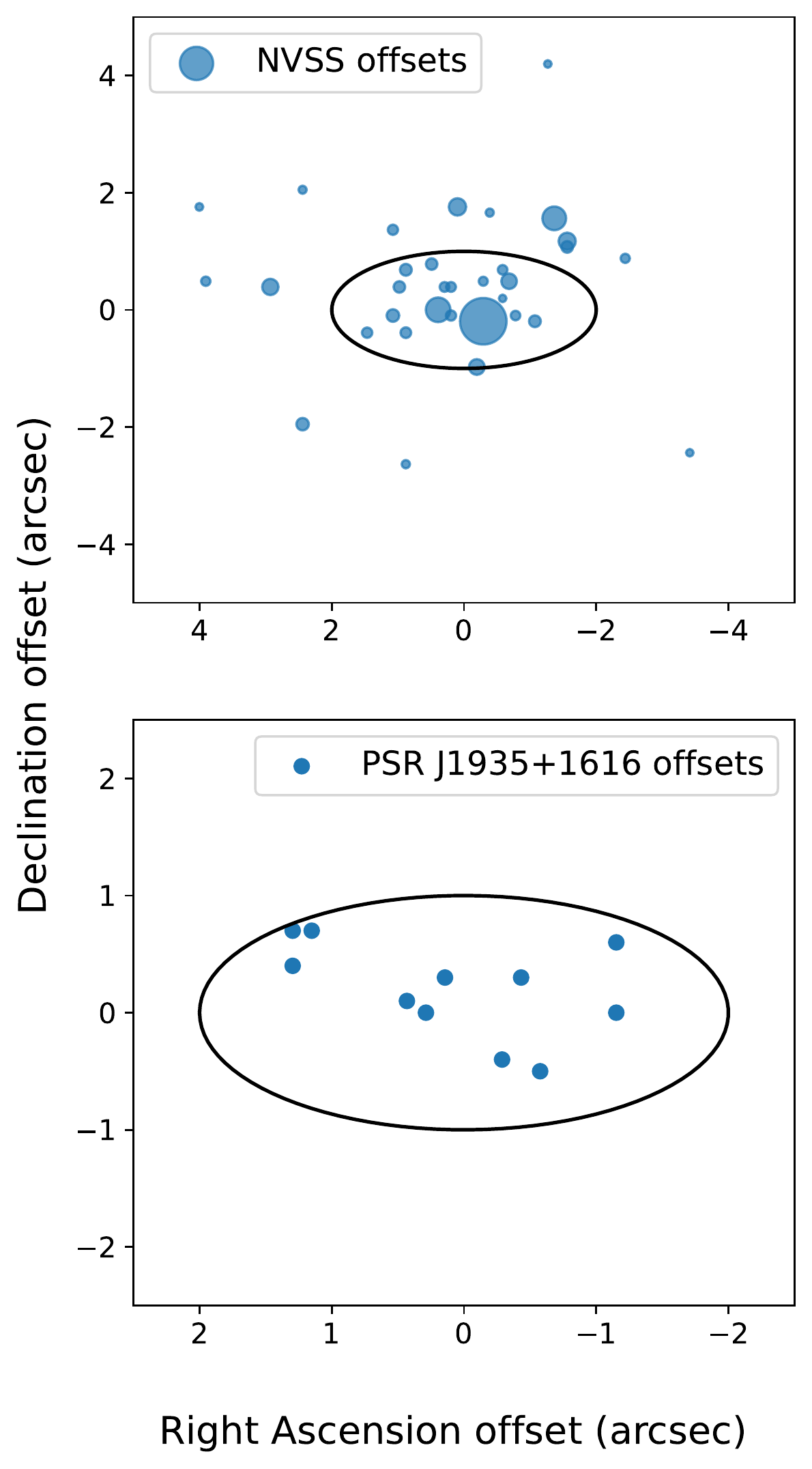}
    \caption{Two different verifications of the accuracy with which DSA-110 recovers the positions of known astronomical sources. Both panels show offsets of sources in DSA-110 data from their true positions, and the approximate 90\% confidence error ellipse quoted for the localization of FRB\,20220912A. \textit{Top:} offsets of sources from cataloged positions in the NRAO VLA Sky Survey \citep[NVSS;][]{ccg+98} in a 10\,min DSA-110 image formed from data taken around Burst 1. Only sources with cataloged flux densities $>50$\,mJy and measured major-axis diameters $<30$\arcsec, within 2\,deg of the pointing center, were considered. The symbol size is proportional to flux density. \textit{Bottom:} Offsets of 11 pulses from PSR\,J1935$+$1616 localized between 2022 October 22 and 2022 October 25 from the true pulsar position, corrected for proper motion. All pulses were localized using exactly the same pipeline as for the bursts from FRB\,20220912A.}
    \label{fig:offsets}
\end{figure}

We were able to derive arcsecond-scale localizations for the two bursts detected with DSA-110 from FRB\,20220912A. The voltage data obtained for each burst were dedispersed with DMs that maximized the S/N, and correlation products in each orthogonal linear polarization were formed and integrated over 8.36\,ms centered on each burst arrival time. These correlation products were written into Common Astronomy Software Applications \citep[\texttt{CASA};][]{casa22} Measurement Sets (MSs), and all further processing was done using \texttt{CASA} version 5.4.1. Bandpass calibrations were obtained using 10-min observations of 3C309.1 on the same sidereal days as the burst detections, and assuming a flat-spectrum point-source model. So-called ``field'' MSs were then generated using 10\,min of visibility data surrounding the arrival times of each burst, fringestopped at an hour angle of 0 at the midpoint of the observation. Sky models based on the NRAO VLA Sky Survey \citep[NVSS;][]{ccg+98} were inserted into the field MSs, including all sources brighter than 30\,mJy of any size. Extended sources were modeled as elliptical Gaussians. After bandpass calibration, the sky models were used to derive complex gain solutions for each antenna from the field MSs. The bandpass and gain calibration tables were then applied to the burst MSs. Wide-field images of the burst MSs 4.5$^{\circ}$ in diameter with 4100 pixels per side were made using the \texttt{wsclean} software \citep{omh+14} to crudely identify the locations of the bursts. The \texttt{CASA} task \texttt{tclean} was then used to make higher resolution images 5\,arcmin and 1000 pixels per side centered on the approximate FRB positions; no deconvolution was applied. In all imaging, natural weighting was used, and baselines shorter than 40\,m were discarded due to spurious correlated power on short baselines. The localizations were derived by fitting two-dimensional Gaussians in the image plane to the central lobe of the resulting point-spread function (PSF) in the burst images.   

The best-fit position for Burst 1 was (R.A. J2000, decl. J2000) = (23:09:04.90, $+$48:42:25.4), and the best-fit position for Burst 2 was (R.A. J2000, decl. J2000) = (23:09:04.88, $+$48:42:25.6).\footnote{\citet{r+22b} issued a correction to the preliminary position for Burst 1 quoted in \citet{r+22a}. The tests on PSR\,J1935$+$1616 described here revealed an error in the calculation of absolute time in the voltage imaging pipeline \citep{r+22b}. This error resulted in a systematic error in the source positions that was not present in the positions measured by the field imaging pipeline. The tests reported here confirm that both pipelines produce consistent astrometry.} Images, with no deconvolution applied, of each burst together with the PSF are shown in Figure~\ref{fig:localizations}. In the widefield \texttt{wsclean} images, the image S/N values for the bursts were 11 and 7, found after 10 image-plane CLEAN iterations and measured far from the sources. As the image of Burst 2 is of somewhat worse quality than that of Burst 1, we adopt the position of Burst 1 as the best-fit DSA-110 position of FRB\,20220912A. 

We verified the astrometric accuracy of our results in four ways. First, we compared images of the primary-beam full-width half-maximum (FWHM) areas made with the entire 2.1\,s voltage data sets, and with the field MSs, and ensured that bright sources were detected at consistent positions. We then checked the efficacy of the gain calibrations derived from the field MSs by checking the positions of bright ($>50$\,mJy) compact ($<30$\arcsec) NVSS sources within the primary-beam FWHM in images of these MSs against NVSS catalog positions. The results are shown in the top panel of Figure~\ref{fig:offsets}. Some offsets in derived positions due to the differing synthesized-beam shapes of the DSA-110 and NVSS are expected for resolved sources, and so this check is best used to search for systematic errors in the gain calibrations rather than quantify the localization accuracy. Instead, we checked the absolute astrometry by forming MSs on and imaging two compact bright sources from the Radio Fundamental Catalog (rfc\_2022c), J2258$+$4937 and J2325$+$4806. These sources are close in declination to FRB\,20220912A and thus transit through similar portions of the DSA-110 primary beam, and are close in right ascension and thus could be calibrated using gains from the field MSs. The derived position offsets were 0.7\arcsec~ and 0.4\arcsec~ respectively from the rfc\_2022c positions on 2022 October 18, and 0.5\arcsec~ and 0.6\arcsec~ respectively on 2022 October 25. We also conducted observations of the bright pulsar J1935$+$1616 over three transits between 2022 October 22 and 2022 October 25, and performed a localization analysis on 11 automatically detected single pulses in exactly the same way as for the FRB\,20220912A burst detections. The resulting offsets from the true position of PSR\,J1935$+$1616 \citep{mht+05}, corrected for proper motion, are shown in the bottom panel of Figure~\ref{fig:offsets}. 

Given these tests, we find no evidence for systematic offsets or effects of non-Gaussian noise. We therefore quote the localization as measured along with a conservative theoretical estimate of the statistical uncertainty for a source with an image S/N of 10. Conservatively, the standard error in each coordinate is given by half the synthesized-beam FWHM \citep{rsm+88}. Each error is scaled by 2.14 \citep[e.g.,][]{ccg+98} to derive diameters of the 90\% confidence error ellipse. We quote an approximate 90\% confidence error ellipse with diameters of 4\arcsec~in right ascension and 2\arcsec~in declination, as shown in Figure~\ref{fig:offsets}.

\section{The host galaxy of FRB\,20220912A} \label{sec:host}

The DSA-110 position for FRB\,20220912A overlaps just a single galaxy, PSO\,J347.2702$+$48.7066 (Figure~\ref{fig:image}). This galaxy, with an $r$-band magnitude of 19.65 \citep{msf+20}, was also noted as a promising host by \citet{mc22}. We therefore consider PSO\,J347.2702$+$48.7066, hereafter PSO\,J347$+$48, as the likely host galaxy of FRB\,20220912A. For example, an analysis of galaxy detections in Pan-STARRS data near the DSA-110 position for FRB\,20220912A with the Probabilistic Association of Transients to their Hosts \citep[PATH;][]{abd+21} with standard priors and a 50\% chance of the host being unseen indicates a 5\% false-association probability.  In this section we present and interpret optical observations of PSO\,J347$+$48. Fundamental parameters that we derive for this galaxy are given in Table~\ref{tab:gal}.

\subsection{Optical observations} \label{sec:optical}

\begin{figure}
    \centering
    \includegraphics[width=0.48\textwidth]{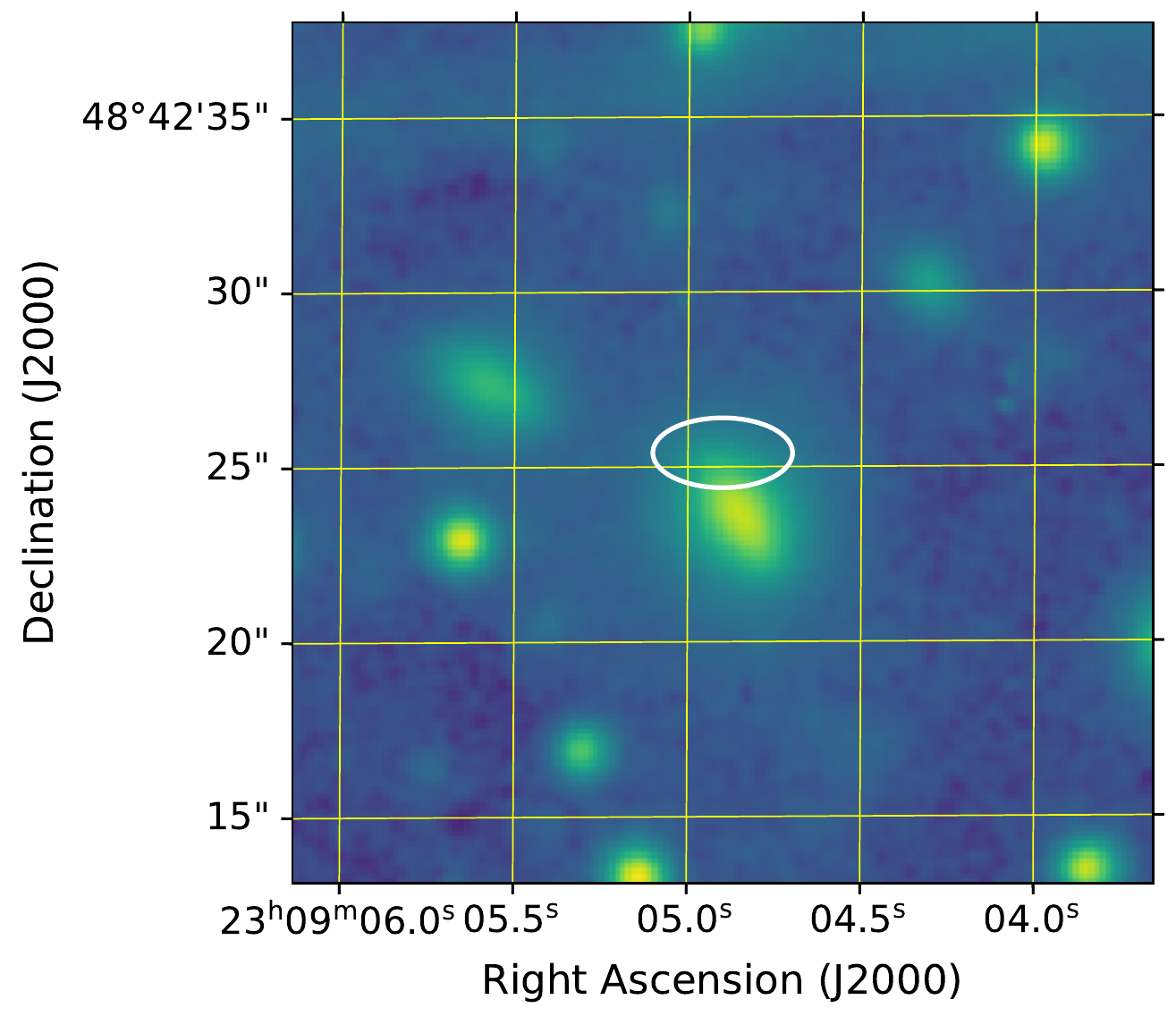}
    \caption{Deep optical $R$-band image obtained with Keck-II/ESI (limiting magnitude $R\sim26$) of the localization region of FRB\,20220912A. The approximate 90\% error ellipse for the DSA-110 localization of FRB\,20220912A is shown in white, to the north of the galaxy PSO\,J347.2702$+$48.7066.}
    \label{fig:image}
\end{figure}

\begin{figure*}
    \centering
    \includegraphics[width=0.85\textwidth]{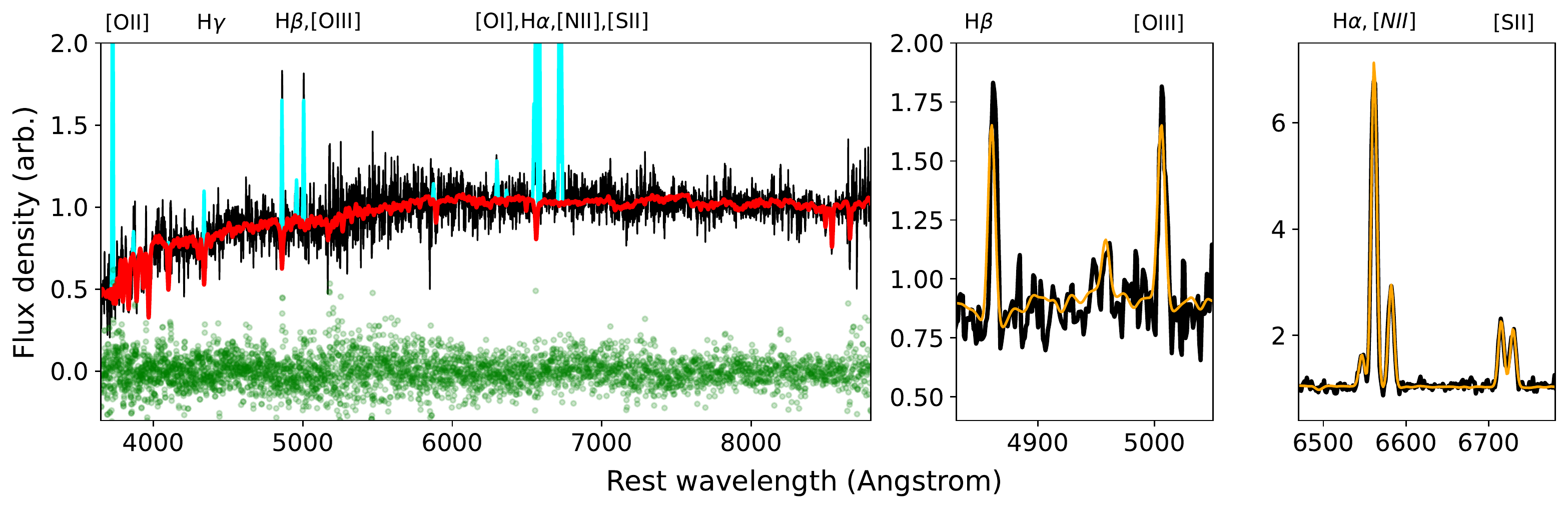}
    \caption{Optical spectrum of the central $\sim1$\,arcsec$^{2}$ of the host galaxy of FRB\,20220912A, PSO\,J347.2702$+$48.7066, obtained with Keck-I/LRIS. The left panel shows the spectrum in black, the \texttt{pPXF} fit to the stellar continuum in red, and the \texttt{pPXF} fit to the nebular emission lines in cyan. The fit residuals are shown as green points. Selected emission lines are indicated. The middle and right panels show zoomed-in views of certain emission lines (black), with the total \texttt{pPXF} fit in orange.}
    \label{fig:spec}
\end{figure*}

We obtained an optical image of PSO\,J347$+$48 in the ``Ellis-$R$'' filter of the Echellette Spectrograph and Imager on the Keck II telescope \citep[Keck-II/ESI;][]{sbe+02}. Four 300\,s exposures were obtained on 2022 October 24 at an airmass of 1.45 in conditions of $0.8$\arcsec~seeing. The data were bias-subtracted, flat-fielded, combined and registered using a custom \texttt{astropy}-based pipeline \citep{astropy22}. An astrometric solution was obtained using the ten nearest unsaturated Gaia Data Release 3 \citep{gaia22} stars to the center of PSO\,J347$+$48. The image is shown in Figure~\ref{fig:image}, and an approximate calibration against Gaia data implies a limiting magnitude of 26. There is no evidence for any object besides PSO\,J347$+$48 within the DSA-110 localization region of FRB\,20220912A. 

We obtained an optical spectrum of PSO\,J347$+$48 using the Low-Resolution Imaging Spectrometer on the Keck I telescope \citep[Keck-I/LRIS;][]{occ+95}. A single 300\,s exposure was obtained on 2022 October 19 in conditions of $0.9$\arcsec~seeing at an airmass of 1.15, using a 1\arcsec~slit at a position angle of 254.7$^{\circ}$. Light was split between the blue and red arms using the D560 dichroic, and dispersed using the 400/3400 grism on the blue arm and the 400/8500 grating on the red arm. The line FWHM was approximately 7\,\AA. The data were processed using standard techniques with the \texttt{lpipe} software \citep{p19}; flux calibration was carried out using observations of the standard star BD$+$28\,4211. A spectrum was extracted from the central square arcsecond of the galaxy; this is shown in Figure~\ref{fig:spec}. 

Several emission lines are evident in the spectrum of PSO\,J347$+$48, at a redshift of $0.0771\pm0.0001$. We measure an H${\rm \alpha}$ line flux of $(1.23\pm0.05)\times10^{-15}$\,erg\,s$^{-1}$\,cm$^{-2}$. The line ratios $\log({\rm [OIII]/H\beta})=0.00\pm0.05$,  $\log({\rm [NII]/H\alpha})=-0.04\pm0.01$, $\log({\rm [SII]/H\alpha})=-0.43\pm0.02$, and $\log({\rm [OI]/H\alpha})=-1.2\pm0.1$ imply a composite BPT classification \citep{kht+03,kgk+06}. A two-dimensional Sersic profile fit to the morphology of PSO\,J347$+$48 in the Keck-II/ESI image, combined with the redshift measurement, indicates an effective radius of $2.2\pm0.1$\,kpc. This is consistent with the $z\sim0$ size-mass relation for late-type galaxies \citep{smw+03}. The H${\rm \alpha}$ line flux implies a star-formation rate in the central square arcsecond of the galaxy of approximately 0.1\,$M_{\odot}$\,yr$^{-1}$, assuming no AGN contribution, and using a conversion factor consistent with \citet{hps+20}. We note that the Keck-I/LRIS two-dimensional spectrum shows evidence for off-nuclear H${\rm \alpha}$ emission. Assuming a disk-like galaxy, the inclination derived from a half-light isophote fit is 51$^{\circ}$. 



\subsection{Host-galaxy modeling} \label{sec:model}

\begin{figure}
    \centering
    \includegraphics[width=0.48\textwidth]{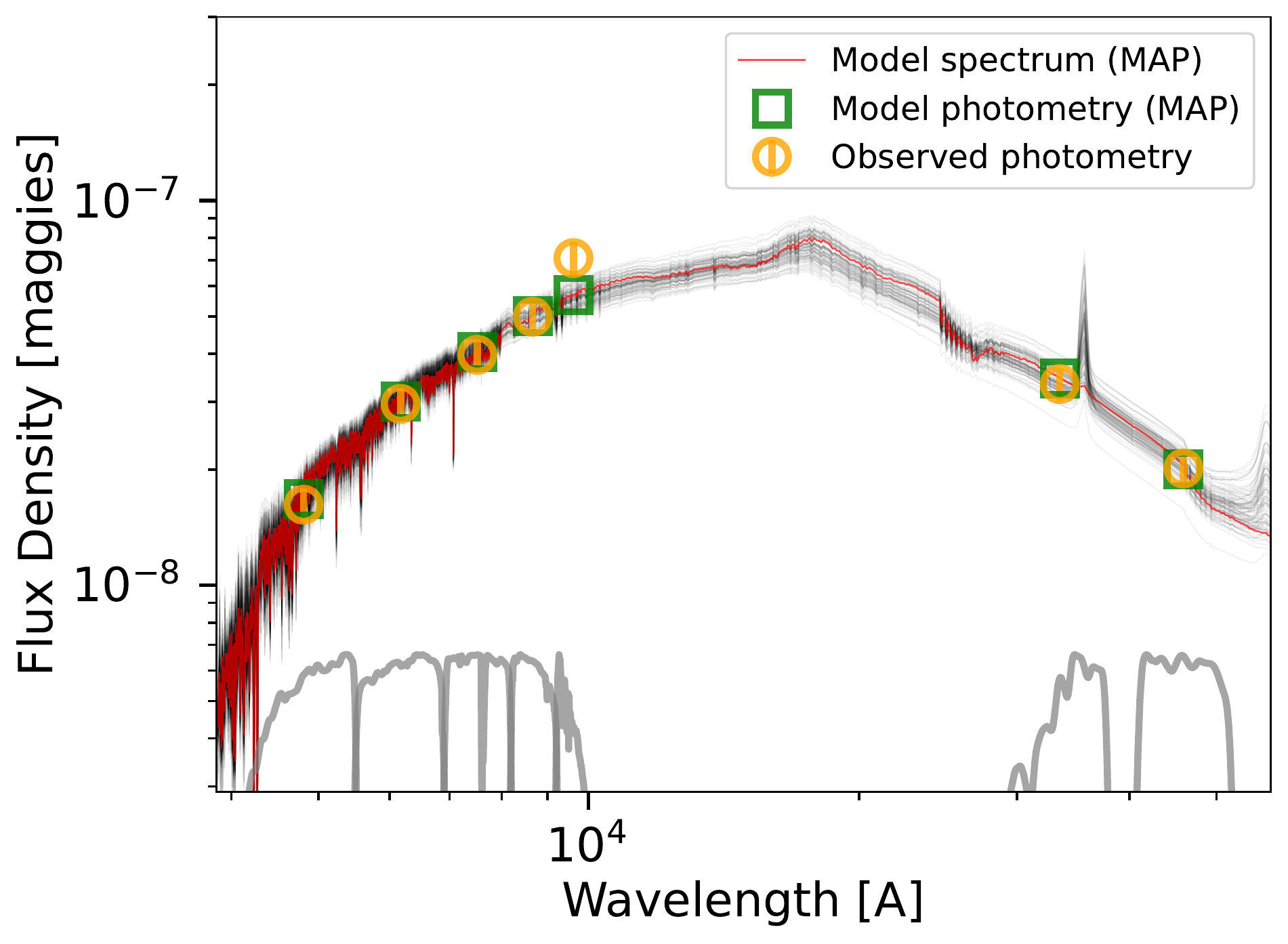}
    \caption{Results of a \texttt{Prospector} fit to the optical/IR SED of the host galaxy of FRB\,20220912A, PSO\,J347.2702$+$48.7066. The SED measurements from archival PanSTARRS and WISE data are shown in orange, together with representative 10\% errors. Filter transmission curves are shown in grey. The maximum aposteriori probability (MAP) model photometry is shown in green, together with the MAP spectrum in red. 100 spectra generated from draws from the posterior distributions on the model parameters are shown in black. }
    \label{fig:sed}
\end{figure}

\begin{deluxetable}{cc}
\tablecaption{Observed and derived parameters of PSO\,J347.2702$+$48.7066. Uncertainties in the last significant figures are given in parentheses.\label{tab:gal}}
\tablehead{
\colhead{Parameter} & \colhead{Value}}
\startdata
Redshift & 0.0771(1) \\
Luminosity distance (Mpc) & 362.4(1) \\
Effective radius (kpc) & 2.2(1) \\
$\log M_{*}$ & 10.0(1) \\
Internal $A_{V}$ & 0.5 \\
$\log Z$ & $-0.8^{-0.3}_{-0.6}$ \\
SFR ($M_{\odot}$\,yr$^{-1}$) & $\gtrsim0.1$ \\
BPT classification & Composite
\enddata
\end{deluxetable}

We performed a fit to the optical spectrum of PSO\,J347$+$48 using the latest iteration of the penalized pixel fitting code \texttt{pPXF} \citep{c22} to jointly fit the stellar continuum and nebular emission. We adopted the default MILES stellar templates and ran the fit using recommended procedures and parameters. The fit results are shown in Figure~\ref{fig:spec}. An internal dust extinction corresponding to $A_{V}=0.5$ was found, and the fit results were also used for the redshift, line-flux and line-ratio measurements given above. 

We then fit archival photometry of PSO\,J347$+$48 from the Pan-STARRS \citep{msf+20} and ALLWISE \citep{cwc+21} source catalogs using the \texttt{Prospector} stellar population synthesis modeling code \citep{jlc+21}. Although the galaxy is marginally detected in 2MASS images, it is not cataloged and we thus did not consider photometry from 2MASS data. ALLWISE detections in bands $w1$ and $w2$ were used, together with all five Pan-STARRS bands. We ran Prospector using recommended techniques and standard priors with a ‘delay-tau’ parametric star-formation history, and sampled from the posterior using
emcee \citep{fhl+13}. We fixed the internal dust attenuation to the \texttt{pPXF} result, and included a model for dust re-radiation in the likelihood. Results for the galaxy stellar mass and metallicity are given in Table~\ref{tab:gal}. The star-formation history was poorly constrained by the fit. The spectral energy distribution of  PSO\,J347$+$48 is shown in Figure~\ref{fig:sed}, together with the results from the Prospector analysis.  

\section{Summary and discussion} \label{sec:summary}

Two bursts detected by the DSA-110 from the repeating source FRB\,20220912A have been used to localize it to (R.A. J2000, decl. J2000) = (23:09:04.90, $+$48:42:25.4). The two bursts have disparate polarization properties. The 90\% confidence error ellipse is $\pm2$\arcsec~ and $\pm1$\arcsec~ in right ascension and declination respectively. The localization is consistent with a single host galaxy, PSO\,J347.2702$+$48.7066, at $z=0.0771\pm0.0001$. The galaxy has a stellar mass of $\log M_{*}=10.0\pm0.1$, and an effective radius of $2.2\pm0.1$\,kpc. Nebular emission lines from the nucleus of this galaxy indicates a ``composite'' BPT classification. 

The DM contributed by the FRB host galaxy along its sightline is low. The extragalactic DM, ${\rm DM}_{\rm cosmic}$, of FRB\,20220912A is approximately 85\,pc\,cm$^{-3}$ \citep{cl02,ymw17}, assuming a modest contribution from the Milky Way halo of 10\,pc\,cm$^{-3}$ \citep{kp20}. If we assume that 70\% of cosmic baryons contribute to the intergalactic DM of FRB\,20220912A, the predicted ${\rm DM}_{\rm cosmic}$ for $z=0.0771$ is 55\,pc\,cm$^{-3}$, implying a host DM contribution of 30\,pc\,cm$^{-3}$. The 90\% confidence lower limit on ${\rm DM}_{\rm cosmic}$ using the model\footnote{The model is implemented in \url{https://github.com/FRBs/FRB}.} for ${\rm DM}_{\rm cosmic}(z)$ presented in \citet{mpm+20} is 32\,pc\,cm$^{-3}$, implying a 90\% confidence upper limit on the host DM of 53\,pc\,cm$^{-3}$. 

The host galaxy of FRB\,20220912A appears unremarkable in the context of FRB hosts, repeating or so far non-repeating \citep{bha+22}. The low host DM of FRB\,20220912A implies a tenuous plasma column towards the source within the host. The small magnitude of the RM reported by \citet{mc22} and confirmed by us, consistent with a negligible extragalactic RM of $\lesssim O(10)$\,rad\,m$^{-2}$, is consistent with the low host DM. The small RM also suggests that the variable polarization properties evident between the bursts we detected from  FRB\,20220912A are an intrinsic effect in the source rather than due to propagation. FRB\,20220912A is a remarkably active repeating FRB, from which we detect pulses with variable morphology and polarization, from a host galaxy much like the hosts of other FRB sources. 

\begin{acknowledgments}

The authors thank staff members of the Owens Valley Radio Observatory and the Caltech radio group, including Kristen Bernasconi, Stephanie Cha-Ramos, Sarah Harnach, Tom Klinefelter, Lori McGraw, Corey Posner, Andres Rizo, Michael Virgin, Scott White, and Thomas Zentmyer. Their tireless efforts were instrumental to the success of the DSA-110. The DSA-110 is supported by the National Science Foundation Mid-Scale Innovations Program in Astronomical Sciences (MSIP) under grant AST-1836018. We acknowledge use of the VLA calibrator manual and the radio fundamental catalog. Some of the data presented herein were obtained at the W. M. Keck Observatory, which is operated as a scientific partnership among the California Institute of Technology, the University of California and the National Aeronautics and Space Administration. The Observatory was made possible by the generous financial support of the W. M. Keck Foundation. 

\end{acknowledgments}

\facility{Keck:I (LRIS), Keck:II (ESI)} 
\software{astropy, CASA, frb, heimdall, lpipe, pPXF, Prospector, wsclean}

\end{document}